%% file: HK-YS.tex
\documentclass[11pt, a4paper]{article}
\usepackage{jheparxiv}
\usepackage{amsmath}
\usepackage{amsfonts}
\usepackage{amssymb}
\usepackage{latexsym}
\usepackage{mathrsfs}
\usepackage{braket}		%Bra Ket Notation
\usepackage{graphicx}
\usepackage{color}
\usepackage{xcolor}
\usepackage{slashed}
\usepackage{twistor}
\usepackage[all]{xy}
\usepackage{tikz-cd}
\usepackage{tikz}
\usepackage{graphicx}
\usepackage{pgfplots}
\pgfplotsset{compat=1.15}

\usetikzlibrary{shapes.geometric, arrows,3d,snakes,intersections}
\usepackage{pgflibrarysnakes}
\usepackage{tikz-3dplot}

\usetikzlibrary{intersections}
\usetikzlibrary{math}
\usetikzlibrary{shapes}
\usetikzlibrary{decorations.markings}
\usetikzlibrary{patterns}
\newcommand{\cY}{\mathcal{Y}}
\newcommand{\cK}{\mathcal{K}}

\renewcommand{\d}{\mathrm{d}}

\newcommand{\qed}{\hfill \ensuremath{\Box}}

\newcommand{\cT}{\mathcal{T}}

\input{figures}

\begin{document}
\begin{center}
\phantom{vv}

\vspace{1cm}
\bigskip

{\Large \bf    Amplitudes at strong coupling as hyperk\"ahler scalars}

\bigskip
{\mbox {\bf Hadleigh Frost\footnotemark, \"Omer G\"urdo\u gan\footnotemark\, \& Lionel Mason$^{1,3}$}%
\setcounter{footnote}{1}
\footnotetext{\, {\texttt{\{\href{mailto:lmason@maths.ox.ac.uk}{lmason}, \href{mailto:Hadleigh.Frost@maths.ox.ac.uk}{Hadleigh.Frost}\}@maths.ox.ac.uk }}}
\setcounter{footnote}{2}
\footnotetext{\,\,\texttt{\href{mailto:o.c.gurdogan@soton.ac.uk}{o.c.gurdogan@soton.ac.uk}}}
}
\bigskip
\end{center}
{\em $^1$The Mathematical Institute,
  University of Oxford, Woodstock Road, OX2 6GG, UK,
  \smallskip\\
  $^2$School of Physics \& Astronomy,
University of Southampton, SO17 1BJ, UK,
\smallskip\\
$^3$Laboratoire de Physique,  \'Ecole Normale Sup\'erieure, Paris 75005, France.

}

\vspace{3cm}
\bigskip
\begin{center}
\vspace{30pt} {\bf Abstract}
\end{center}

\noindent Alday \& Maldacena conjectured  an equivalence between string amplitudes in AdS$_5 \times S^5$ and null polygonal Wilson loops  in  planar $\cN=4$ super-Yang-Mills (SYM).  At strong coupling this identifies SYM amplitudes with areas  of minimal surfaces in AdS. For minimal surfaces in AdS$_3$, we find that the nontrivial part of these amplitudes, the \emph{remainder function}, satisfies an integrable system of nonlinear differential equations, and we  give its Lax form. The result follows from a new perspective on `Y-systems', which defines a new psuedo-hyperk\"ahler structure \emph{directly} on the space of kinematic data, via a natural twistor space defined by the Y-system equations. The remainder function is the (pseudo-)K\"ahler scalar for this geometry. This connection to pseudo-hyperk\"ahler geometry and its twistor theory provides a new ingredient for extending recent conjectures for non-perturbative amplitudes using structures arising at strong coupling.

%%%%%%%%%%%%%%%%%%%%%%%%%%%%%%%%%%%%%%%%
%%%%%%%%%%%%%%%%%%%%%%%%%%%%%%%%%%%%%%%%
\newpage
\phantom{vv}
\vspace{1cm}
\hrule
\tableofcontents

\bigskip
\medskip

\hrule
%\newpage
\section{Introduction}
The  spaces of kinematic data $\cK_n$ on which $\cN=4$ super Yang-Mills amplitudes are defined has a rich combinatorial that has been a fertile ground for advancing the understanding of scattering amplitudes \cite{Golden:2013xva, Arkani-Hamed:2012zlh,Arkani-Hamed:2013jha}. This paper discovers new geometric structures that arise from the analysis of the amplitude at strong coupling \cite{Alday:2009yn,Alday:2010vh} and that complements the combinatorial cluster variety and positivity structures that arise at weak coupling. The successes of the bootstrap of \cite{Dixon:2011pw, Caron-Huot:2020bkp} hinges on the cluster geometry of kinematic spaces $\cK_n$ but more recently features from strong coupling have played an instrumental role in generating the nonperturbative conjectures of \cite{Basso:2020xts,Basso:2022ruw}.  Our new integrable geometric structures encode the full structure of the strong coupling amplitude and will provide foundations for further advances in this direction.

Alday-Maldacena \cite{Alday:2007hr,Alday:2007he} conjectured a 3-way correspondence between  planar  amplitudes, $\cA$; planar null-polygonal Wilson-loops, $\langle \cW_\gamma \rangle$, both for $\cN=4$ super-Yang-Mills; and   type IIB string amplitudes in AdS$_5\times S^5$
\begin{equation}
\cA =\langle \cW_\gamma \rangle = \int_{\p\Sigma=\gamma} \cD[\Sigma \subset AdS_5\times S^5] \; \e^{- \frac{1}{\alpha'}
S_{\mathrm{string}}} \, .
\end{equation}
Here $\gamma$ is a null polygon made up from the null momenta in the amplitude. The $\alpha'$ string parameter  is related to the 't Hooft coupling, $\lambda$, by $R^2_{AdS} / \alpha' = \sqrt{\lambda}$. The first equality has been proved\footnote{The tree-level MHV amplitude is removed in the definition of $\cA$.} in perturbation theory  \cite{Mason:2010yk,Caron-Huot:2010ryg}.  The second equality is a conjecture arising from the AdS/CFT correspondence. It has only been systematically investigated at strong coupling as $\lambda\rightarrow \infty$ (and $\alpha'\rightarrow 0$), where the semi-classical  approximation for the string gives
\begin{equation}
\langle \cW_\gamma\rangle \sim \e^{-\mathrm{Area}(\Sigma)/\alpha'}\, ,
\end{equation}
where $\mathrm{Area}(\Sigma)$ is the area of the minimal surface, $\Sigma$, bounded by $\gamma$. Like the Wilson-loop, $\langle \cW_\gamma\rangle$, the area of the minimal surface $\Sigma$ is divergent at its cusps  where it meets the boundary at infinity. These correspond precisely to the infrared divergences of the amplitude. These divergences can be removed compatibly with all three interpretations leading to a regularized area or remainder function, $R(\gamma)$, which is our main object of study.

Alday-Maldacena reformulate minimal surfaces in AdS as a Hitchin system and express the area as the Hamiltonian for a certain circle action on the kinematic data \cite{Alday:2009yn}.  Hitchin moduli spaces are often hyperk\"ahler \cite{ Hitchin:1986vp, Hitchin:1988df} but discrete symmetries are imposed to give minimal surfaces so that standard results (from, e.g., \cite{Biquard:2004,Gaiotto:2009hg}) do not apply, and our space  is not expected to be hyperk\"ahler from these arguments, see \S3.3 of \cite{Alday:2009yn}. 
However, we will show that these smaller moduli spaces are often \emph{pseudo-hyperk\"ahler}, i.e., the analogue of hyperk\"ahler appropriate to metrics of split signature.

%\begin{figure}
%    \centering
%\cylinderzigzag
%        \caption{Minimal surface in AdS$_3$ with boundary on null polygon at infinity }
%    \label{fig:enter-label}
%\end{figure}

Although the main structures we use are available for full kinematics, in this paper we work with momenta and Wilson loop lying in $1+1$ dimensions with the spanning minimal surface living in AdS$_3$.  Although this might seem  restrictive, in practice the extension to full kinematics is well known to be a straightforward  but elaborate extension to larger cluster varieties \cite{Alday:2010ku,Gao:2013dza} where the extra complexity will obscure the essential ideas.  We therefore postpone this discussion. We prove here that the regularized area of this surface is a K\"ahler scalar for a pseudo-hyperk\"ahler structure on $\cK_n$, when it has $4k$ dimensions.  We do this by using the $Y$-system \cite{Alday:2010vh} to define a twistor space for $\cK_n$, analogous to the twistor spaces for full Hitchin moduli spaces; we expect this novel connection between  $Y$-systems and twistor constructions to be of much wider applicability.  Then we derive a system of integrable equations satisfied by the regularized area, which can be used to solve for the area. In \S\ref{kinematics} we introduce the kinematic space $\cK_n$ and both its cluster and associated Poisson or symplectic structure. In \S\ref{Y-sys-twis} we recall the $Y$-system \cite{Alday:2010vh} and explain how it defines a twistor space for $\cK$. In \S\ref{DEs} we find the hyperk\"ahler structure explicitly and show the regularized area is a K\"ahler scalar for a split signature analogue of a hyperk\"ahler metric that satisfies an integrable system of generalized Plebanski equations. Finally in \S\ref{discuss} we  mention a number of checks and further developments including applications to amplitudes at finite coupling.

%\newpage
\section{ The spaces of kinematic data in $1+1$-dimensions}\label{kinematics}
Our kinematic space $\cK_n$ here will be  the moduli space of $2n-sided$ null polygonal Wilson loops in $1+1$-dimensions.  Such  Wilson loops are given by a set of ordered null momenta (the `edges' of the loop) that sum to zero (so that the loop closes). Take null coordinates $(X^+,X^-)$ on Minkowski space with metric
\begin{equation}
\d s^2=2\d X^+\d X^-.
\end{equation}
The edges of a polygonal Wilson-loop alternate between lines of constant $X^+$ and constant $X^-$. The kinematic data for a $2n$-sided Wilson loop in AdS$_3$ is therefore given by two cyclically ordered sets of real numbers $\{X_i^+\},\{X^-_i\}$, with $i=0,\ldots ,n-1$. Vertices of the polygon are given by the points $(X^+_i,X^-_{i-1})$,  $(X^+_i,X^-_{i})$ then $(X^+_{i+1},X^-_i)$ and so on.  Conformal invariance means that our functions of these parameters should be invariant under M\"obius transformations on the $X^+_i$ and  $X^-_i$ separately. Thus the space of kinematic data $\cK_n$ is
\begin{equation}
\cK_n = \cM^\R_{0,n}\times \cM^\R_{0,n}\, .
\end{equation}
where   
\begin{equation}
\cM^\R_{0,n} =\{X^\pm_i, i=1,\ldots ,n\}/PSL_2\, .
\end{equation}
is the moduli space of $n$ points on $\RP^1$ modulo M\"obius transformations.

\begin{figure}
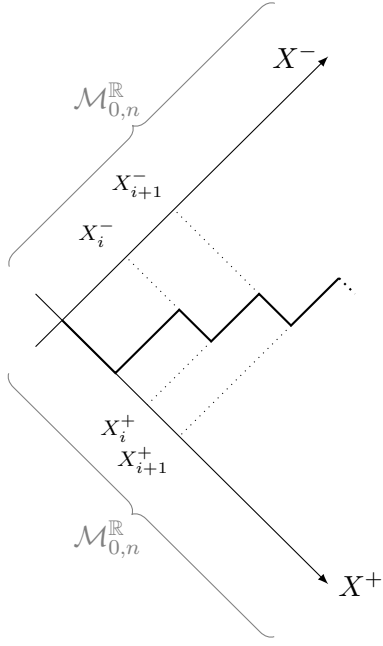

% \hskip-0.5cm	
\begin{center}
  \zigzaganan
 %\hskip3cm
%\vspace{-0.4cm}
\end{center}
\caption{{\it Null coordinates on $\M^{1,1}$.}}
\label{can}
\end{figure} 

M\"obius invariant functions on $\cK$ are functions of cross-ratios
\begin{equation}
(ij|kl)^\pm = \frac{(X^\pm_i - X^\pm_j)(X^\pm_k-X^\pm_l)}{(X^\pm_i - X^\pm_l)(X^\pm_j-X^\pm_k)}.
\end{equation}
Our charts for
$\mathcal{K}$ are sets of cross-ratios called \emph{clusters} \cite{Fock:2009}. A cluster is specified by choosing a \emph{triangulation} of the $n$-gon. For a fixed triangulation, the chords are indexed by $s=1,\ldots ,n-3$. Such a chord connects say vertex $i$ to $k$ making the diagonal $i-k$ of some quadrilateral $(i,j,k,l)$ formed by two triangles.  To this chord  associate the coordinate
\begin{equation}
    \chi_s^\pm = (il|kj)^\pm.
\end{equation}
The set of these cross ratios, $\{\chi_s^\pm\}$ define a cluster of coordinates on $\cK_n$.

\begin{figure}
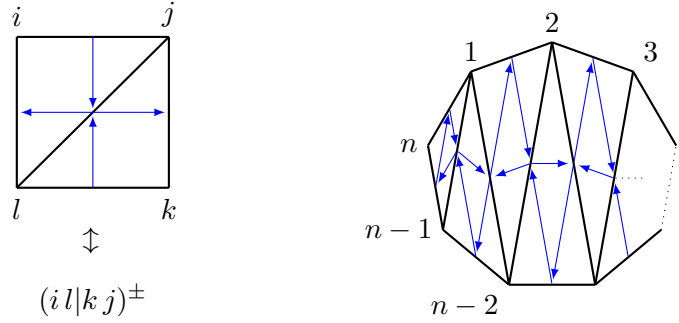
\label{zigzag}
\begin{center}
\exxodd \qquad  \zigzagcluster 
\end{center}
\caption{Left: The correspondence between chords of a 
  triangulation and cross ratios.  Right: The zig-zag triangulation of the
  polygon.}
\end{figure}

Different choices of clusters of coordinates are related by \emph{mutation relations}. Flipping a chord $s$ inside a quadrilateral that it is a diagonal of, gives a new chord, $s'$, and the new cross-ratios are related to the old ones by
\begin{equation}\label{eq:mutrule}
\mu(\chi_{s}) = \chi_s^{-1}, \qquad \mu(\chi_t) = \chi_t (1+\chi_s^{\epsilon_{st}})^{\epsilon_{st}}.
\end{equation}
The only cross-ratios that change are those sharing a triangle with $s$ in the triangulation. 

Finally, there is a 2-form on $\cK_n$ that is symplectic when $\cK_n$ is even dimensional ($n$ odd). Fixing a triangulation of the $n$-gon as above, define an antisymmetric matrix,
$\epsilon_{ss'}$  where for chords $s$ and $s'$, write
$\epsilon_{ss'} = 0$ if the two chords do not share a triangle. If they share a triangle, write $\epsilon_{ss'}=1$ if $s'$ is clockwise of $s$, or write $\epsilon_{s's}=-1$ is $s'$ is counter-clockwise of $s$. On each copy of $\cM^\R_{0,n}$ define
\begin{equation}\label{eq:symone}
\omega^\pm = \sum_{i,j} \epsilon_{ij}\, \d\log \chi_i^\pm \wedge \d \log \chi_j^\pm\, .
\end{equation}

We fix, for example, the zig-zag triangulation of Figure \ref{zigzag}, as in \cite{Alday:2010vh}. This gives cross-ratios 
\begin{equation}
\chi_s^\pm = \left\{\begin{matrix} (s-1,s|-s-1,-s)^\pm & \text{$s$ odd,} \\ (s-1,s|-s,-s+1)^\pm & \text{$s$ even,} \end{matrix} \right. \qquad s=1,...,n-3\label{zig-zag}
\end{equation}
where we specify vertices of the polygon mod $n$. The matrix $\epsilon_{ss'}$ is given by $\epsilon_{s,s+1}=1$,  $s$ odd,  and $\epsilon_{s,s+1}=-1$ for $s$ even giving   
\begin{equation}
\omega^\pm = \sum \d\log \chi_{2i}^\pm \wedge ( \d\log \chi_{2i-1}^\pm - \d\log \chi_{2i+1}^\pm ).
\end{equation} 

The symplectic structures is invariant under mutations. \cite{Fock:2009} After one mutation using \eqref{eq:mutrule}, the symplectic 2-form becomes
\begin{equation}\label{eq:symmut}
\omega^\pm=\mu(\omega^\pm) := \sum_{i,j} \tilde{\epsilon}_{ij} \d\log \mu(\chi_i^\pm) \wedge \d \log \mu(\chi_j^\pm),
\end{equation}
where $\mu(\chi_i^\pm)$ are the new cross-ratios, and $\tilde{\epsilon}_{ij}$ is the matrix of the mutated triangulation. Thus, a series of mutations preserves $\omega^\pm$ so $\omega^\pm$ is independent of the choice of cluster.

\section{From the Y-system to the  twistor space}\label{Y-sys-twis}
%\begin{figure}
%\begin{center}
%\begin{tikzpicture}
%     \tikzcuboidset{hidden edges/.style={dashed,draw=white},all faces/.style={fill=gray}, edges/.style={draw=white,thick}}
%      \pic at (10,1,0) {cuboid=2--1--3};
%\node (10) at (13.7,2.5) {$\stackrel{p}{\rightarrow }$};
%
%\node (12) at  (12.4,0.0) {$\cK \ni (\chi_s^+,\chi^-_s)$};
%\node (11) at (11.5,0.5) {${\downarrow }$};
%\draw[thick] (14.2,2.7) -- (14.2,1.7);
%\node (14) at  (14,1.4) {$\zeta\in\CP^1$};
%\node (15) at  (10.5,2.7) {$\cT_n$};
%   \end{tikzpicture}  
%\hskip-1cm
%\end{center}
%\caption{The twistor space, $\cT_n$, is fibred over $\CP^1$.}
%\end{figure}

%The Y-system can be used to construct a twistor space, which will we use in Section \ref{DEs} to prove new results about the remainder function.

 The Y-system of \cite{Alday:2010vh} is based on the zig-zag cluster coordinates $(\chi_s^+,\chi_s^-)$ on $\cK_n$ (equation \eqref{zig-zag}).  The Y-system associated to this cluster  consists of a family of functions 
\begin{equation}
\cY_s=\cY_s(\chi^+_r,\chi^-_r,\zeta):\cK_n \times \CP^1\longrightarrow \C,\qquad s=1,\ldots,n-3
\end{equation}
that are complex analytic in the \emph{spectral parameter} $\zeta\in \CP^1$ fixed by the following four conditions. First,  at  $\zeta=1, i$ we have:
\begin{equation}\label{eq:special}
\cY_s(1)=\chi_s^+\, , \qquad \cY_s(i)=\chi^-_s\, . 
\end{equation}
Second, the $\cY_s$, are holomorphic except for branching singularities at $\zeta=0$ and $\zeta=\infty$, with a branch cut along $\R^-$.  Thirdly, we require exponential asymptotics at the singularities:
\begin{equation}
\log \cY_s \simeq Z_s \zeta^{-1} + \ldots\, , \quad \mbox{ as  } \zeta\rightarrow 0\, , \qquad  \log\cY_s \simeq \bar{Z}_s \zeta + \ldots, \quad \mbox{ as } \zeta\rightarrow\infty    \label{Asymptotics}
\end{equation}
 for some $Z_s(\chi_r^+,\chi_r^-)$. 
Finally, we define the analytic continuation $\cY_s$ across $\R^-$. Writing $\cY^{++}_s(\zeta)=\cY (\e^{i\pi} \zeta)$,  the analytic continuation for the zigzag cluster is given by
\begin{align}
\cY^{++}_{2k+1}\cY_{2k+1}&=(1+\cY_{2k+2})(1+\cY_{2k})\, ,\nonumber \\
\cY^{++}_{2k} \cY_{2k}&=(1+\cY^{++}_{2k+1})(1+\cY^{++}_{2k-1})\, .\label{monodromies}
\end{align}
These relations are defined so that analytic continuation implements a series of mutation relations on the $\chi_s^\pm$, \eqref{eq:mutrule}, that \emph{rotate} the initial triangulation by $2\pi / n$. In other words, $\cY_s^{++}(1)$ and $\cY_s^{++}(i)$ are the cross-ratios obtained by performing this series of mutations. %This rotation is achieved by applying $2(n-3)$ mutations.

The $\cY_s$ are uniquely determined by these conditions as can be seen by iteration of the equivalent integral equations of the Thermodynamic Bethe Ans\"atze (TBA) described in \cite{Alday:2009dv, Alday:2010vh}. 

We now define the \emph{twistor space} to be\footnote{ We suppress the proper description of the real slice $|\zeta|=1$ as treated in the analogous case of \cite{Lebrun:2007}.}  $\cT_n=\cK_n\times \CP^1$  as a smooth manifold. The $\cY_s$-functions define holomorphic coordinates on $\cT_n$, making it a complex manifold.

\begin{propn}\label{prop:twist}
 $\cT_{n}$ is a complex  $n-2$-manifold with local holomorphic coordinates $(\cY_s, \zeta)$ and holomorphic projection: $p:\cT_{n}\rightarrow \CP^1$. There is a family of symplectic 2-forms $\Sigma(\zeta)$ on the fibres of $p$. For odd $n$, $\Sigma(\zeta)$ is non-degenerate. Moreover, $(\cT_n,\Sigma(\zeta))$ is invariant under the holomorphic circle action
 \begin{equation}\label{eq:circa}
 (\cY_s,\zeta)\longrightarrow (\cY_s,\e^{i\theta}\zeta).
 \end{equation}
 Finally, there is an anti-holomorphic  involution on $\cT_n$,
 \begin{equation}
     (\cY_s,\zeta)\longrightarrow (\overline{\cY}_s,1/\bar\zeta),
 \end{equation}   
so  that the $\cY_s$ are real on the unit circle $|\zeta|=1$.
\end{propn}

\noindent
{\bf Proof:} Construct $\mathcal{T}_n$ by gluing together holomorphic coordinate patches 
\begin{equation}
U= \{-\pi < \arg \zeta <\pi\}\, , \quad \mbox{ and }\quad U^{++}=\{0<\arg\zeta <2\pi\}    \, ,
\end{equation}
glued by  $\zeta \mapsto e^{i\pi}\zeta$. The $\cY_s$ functions are holomorphic coordinates on $U$ and the $\cY^{++}_s$  on $U^{++}$. These two sets of holomorphic coordinates are glued together on   $U\cap U^{++}$ by the Y-system equations \eqref{monodromies}.

For fixed $\zeta$, define 
 \begin{equation}
\Sigma(\zeta) := \sum \epsilon_{ij} \d  \log \cY_i \wedge d  \log \cY_j. \label{Sigma-def}
\end{equation}
We claim that this closed 2-form is preserved by mutations.\footnote{A direct proof is to define a generating function for \eqref{monodromies}. A conceptually interesting proof is to apply a series of `mutations on sinks', as in the proof of Zamolodchikov's periodicity conjecture given, for example, by Theorem 8.8 of \cite{fomin2007cluster}; see \cite{williams2013cluster} for a review.} In particular, $\Sigma^{++}(\zeta) \equiv \Sigma(e^{i\pi}\zeta) = \Sigma(\zeta)$. So $\Sigma(\zeta)$ is defined for all $\zeta$, except for $\zeta=0$ and $\zeta=\infty$. Moreover, $\cY_s$ are invariant under the circle symmetry, so $\Sigma$ is likewise circle invariant.

Finally, the functions $\overline{\cY}_s(1/\bar\zeta)$ have the same analytic properties and special values as the functions $\cY_s(\zeta)$, and  satisfy the same Y-system equations as $\cY_s(\zeta)$. But the $\cY_s$ functions are unique, so that $\overline{\cY}_s(1/\bar\zeta) = \cY_s(\zeta)$.
\qed

\section{Integrable system for the remainder function}\label{DEs}
The \emph{remainder function}, $R(\chi^+_r,\chi^-_s)$, is the nontrivial part of the regularised area of the minimal surface in AdS$_3$. Here we define $R$ to be the Hamiltonian for the circle-action, \eqref{eq:circa} (following Sec. 3 of \cite{Alday:2009yn}). Our main result is that $R$ satisfies an integrable system on kinematic space, $\cK_n$. In fact, $R$ defines a pseudo-hyperk\"ahler structure on $\cK_n$.

A \emph{pseudo-hyperk\"ahler} structure is the analogue of a hyperk\"ahler structure, \cite{Berger:1955,Hitchin:1986ea} but with a split-signature metric, as in \cite{Dunajski:2020qhh}.  This structure consists of a split-signature metric, together with three symplectic  2-forms, $\omega^\pm$ and $\Omega$. The 2-forms are compatible with the metric if they satisfy the \emph{pseudo-quaternion} relations. Using the metric to raise the indices of the 2-forms to obtain tensors, $J^\pm$ and $\Omega^\sharp$, the pseudo-quaternion relations are: $(J^+)^2=(\Omega^\sharp)^2=1$, $(J^-)^2=-1$ and $J^-J^+=-J^+J^-=\Omega^\sharp$, $\{J^\pm,\Omega^\sharp\}=0$.

\begin{propn}\label{prop:kahler}
For $n$ odd, $\cK_n$ is pseudo-hyperk\"ahler, with split-signature metric
\begin{equation}
ds^2:=R^{rs} \d x^+_r\d x^-_s, 
\end{equation}
and the three symplectic 2-forms
\begin{equation}
\quad \omega^\pm=\epsilon^{rs}\d x_r^+\wedge\d x^+_s\pm \epsilon^{rs}\d x_r^-\wedge\d x^-_s\, , \qquad \Omega=R^{rs}\d x_r^+\wedge\d x^-_s \,.
\end{equation}
\end{propn}

\noindent
Proposition \ref{prop:kahler} can be shown as a consequence of the following result, which shows that $R$ satisfies the analogue of the \emph{first Plebanski equation}.

\begin{propn}\label{prop:pleb}
For $n$ odd, the remainder function satisfies
\begin{equation}\label{Pleb-R0}
R^{pq}R^{rs}\epsilon_{pr}=\epsilon^{qs},
\end{equation}
together with circle invariance (equation \eqref{circle0}, below). Here $\epsilon_{pq}\epsilon^{qr} = \delta_p^r$ and $R^{pq}$ is the Hessian:
\begin{equation}
R^{rs}=\frac{\p^2 R}{\p x^+_r\p x^-_s}, 
\end{equation}
with $x_p^\pm \equiv \log \chi_p^\pm$. This is an integrable system with Lax system $\{\cL_r, \tilde V\}$, where
\begin{equation}
\cL_r:= (\zeta^2-1)\frac{\p}{\p x^+_r} + (\zeta^2+1)i R^{rs}\frac{\p}{\p x^-_s}\, ,\qquad \widetilde V:=\epsilon_{rs}\left(\frac{\p R}{\p x^+_s}\frac{\p}{\p x^+_r}+\frac{\p R}{\p x^-_s}\frac{\p}{\p x^-_r}\right)+ i\zeta\frac{\p}{\p\zeta}.
\end{equation}
\end{propn}

A \emph{pseudo-hyperk\"ahler} structure is the analogue of a hyperk\"ahler structure, but with a split-signature metric.

\begin{propn}\label{prop:kahler}
For $n$ odd, $\cK_n$ is pseudo-hyperk\"ahler, with split-signature metric
\begin{equation}
ds^2:=R^{rs} \d x^+_r\d x^-_s, 
\end{equation}
and the three symplectic 2-forms
\begin{equation}
\quad \omega^\pm=\epsilon^{rs}\d x_r^+\wedge\d x^+_s\pm \epsilon^{rs}\d x_r^+\wedge\d x^+_s\, , \qquad \Omega=R^{rs}\d x_r^+\wedge\d x^-_s \,.
\end{equation}
\end{propn}

Proposition \ref{prop:kahler} follows from the proof of Proposition \ref{prop:pleb}.

\noindent
{\bf Proof.} Consider a Laurent series expansion of $\Sigma(\zeta)$ in $\zeta$. Equation \eqref{eq:special} implies that $\Sigma(\zeta)$ has special values
\begin{equation}\label{eq:sigone}
\Sigma(1) = \sum \epsilon_{ij} \d x_i^+ \wedge \d x_j^+,\qquad\text{and}\qquad \Sigma(i) = \sum \epsilon_{ij}\d x_i^- \wedge \d x_j^-.
\end{equation}
Moreover, writing $y_s  \equiv \log \cY_s$, $y_s \sim 1/\zeta$ at $0$ and $y_s \sim \zeta$ at $\infty$. So $\Sigma(\zeta)$ has \emph{double poles} at $\zeta=0$ and $\infty$. By Proposition \ref{prop:twist},
\begin{equation}\label{eq:sigtwo}
\Sigma(-\zeta)=\Sigma(\zeta),
\end{equation}
so that the Laurent series only contains even powers of $\zeta$. Equations \eqref{eq:sigone} and \eqref{eq:sigtwo} imply that the Laurent expansion is
\begin{equation}\label{eq:sigexp}
\Sigma(\zeta)= \frac{(\zeta^2+1)^2}{4\zeta^2}\Sigma(1)-\frac{(\zeta^2-1)^2}{4\zeta^2}\Sigma(i)+\frac{(\zeta^4-1)}{4\zeta^2}i\Omega\, .
\end{equation} 
for some $\zeta$-independent closed 2-form $\Omega$. 
%Grouping terms gives
%\begin{equation}\label{eq:sigA}
%\Sigma(\zeta) = \frac{1}{\zeta^2} A^{(-2)} + A^{(0)} + \zeta^2 A^{(2)},
%\end{equation}
%where
%\begin{align}\label{eq:A}
%A^{(\pm 2)} &= \frac{1}{4} \Sigma(1) - \frac{1}{4} \Sigma(i) \mp \frac{i}{4} \Omega
%,\qquad
%A^{(0)} = \frac{1}{2} \Sigma(1) + \frac{1}{2} \Sigma(i).
%\\A^{(2)} &= \frac{1}{4} \Sigma(1) - \frac{1}{4} \Sigma(i) - \frac{i}{4} J^{rs} d  x^+_r\wedge \d x^-_s.
%\end{align}

Since $\Sigma(\zeta)$ is non-degenerate with rank $n-3$,
\begin{equation}
(\Sigma(\zeta))^{(n-1)/2}=0\, . \label{rank-cond}
\end{equation}
%In particular, taking a Laurent expansion of \eqref{rank-cond}, the leading coefficients (multiplying $\zeta^{n-1}$ and $\zeta^{-(n-1)}$) must vanish: $\left(A^{(\pm 2)}\right)^{(n-1)/2} = 0$. 
Taking the derivative with respect to $\zeta$ at $\zeta=1, i$ gives respectively
%components of these coefficients (those with at most two $dx^{+}$'s or at most two $dx^{-}$'s) gives the following two vanishing statements, after a small manipulation:
\begin{align}
(\Sigma(1))^{(n-3)/2}\wedge \Omega &= 0\, , \qquad
(\Sigma(i))^{(n-3)/2}\wedge \Omega  = 0\,.
\end{align}
Since $ (\Sigma(1))^{(n-3)/2}\neq 0$  is of top degree  in the  $dx^+_r$ alone,  the first implies that $\Omega$ is at least linear in the $\d x_r^+$ and the second  that it is at least linear in  $\d x_s^-$, so:
\begin{equation}\label{eq:omega}
\Omega=\frac{1}{4} J^{rs} \d  x^+_r\wedge \d x^-_s
\end{equation}
for some $J^{rs}(x^+,x^-)$. But $\Omega$ is closed, so
\begin{equation}
J^{rs}=\frac{\p^2 J}{\p  x^+_{r} \p x^-_{s}},
\end{equation}
for some scalar  $J(x^+,x^-)$. Again using the vanishing of the leading terms in the expansion of \eqref{rank-cond}, and extracting components with exactly two $\d x^+$'s (or two $d x^-$'s) gives
\begin{equation}\label{eq:J}
\epsilon_{rr'} J^{rs}J^{r's'}=\epsilon^{ss'}\,.
\end{equation}

There is a Lax system for \eqref{eq:J}. Since $\Sigma(\zeta)$ has rank $n-3$, it has $n-3$ null directions. It can be checked using \eqref{eq:sigexp} and \eqref{eq:omega} that  these null directions are  spanned by 
\begin{equation}
\cL_r \equiv (\zeta^2-1)\frac{\p}{\p x^+_r} + (\zeta^2+1)i J^{rs}\frac{\p}{\p x^-_s}\, ,\label{eq:cLr}
\end{equation}
for $r=1,...,n-3$. Since $\cL_r \lrcorner \Sigma(\zeta)=0$, it follows that the $\cL_r$ are involutive. In fact, using \eqref{eq:J} and \eqref{eq:cLr}, these vector fields commute: $[\cL_r,\cL_s] = 0$. Finally, since $\epsilon_{ij}$ is non-degenerate, and since the $\cL_r$ span the kernel of $\Sigma(\zeta)$, it follows that these vector fields annihilate the Y-functions: $\cL_r \cY_s(\zeta)=0$. This can be used to solve for the Y-functions. Near $\zeta = 1,i$, we find
\begin{equation}
\log \cY_r(\zeta)= x_r^\pm +(\zeta^2\mp 1)\epsilon_{rs} i\frac{\p J}{\p x^\pm_s}+ O((\zeta^2\mp 1)^2)\, ,
\end{equation}
and the Lax system further determines all higher order terms.

Let $V$ be the vector field on $\cK_n$ generating the circle action that rotates the phases of the $Z_r$ of \eqref{Asymptotics}.  The \emph{remainder function}, $R$, was shown in \cite{Alday:2009yn} to be the Hamiltonian for $V$ with respect to the symplectic form $\Sigma(1) +\Sigma(i)$ obtained as the coefficient of  $\zeta^0$ in $\Sigma(\zeta)$. 
The Hamiltonian equation is $dR = V \lrcorner (\Sigma(1)+\Sigma(i))$. This implies that
\begin{equation}
V=\epsilon_{rs}\left(\frac{\p R}{\p x^+_s}\frac{\p}{\p x^+_r}+\frac{\p R}{\p x^-_s}\frac{\p}{\p x^-_r}\right)\, . \label{Ham+-}
\end{equation}
We can relate this to $J$ as follows.  The   lift $\tilde V$ of $V$ to $\cT_n$ acts on $\zeta$ by $\tilde{V}(\zeta) = i \zeta$ and so is given by $\tilde{V} = V + i \zeta \partial_\zeta$. 
On $\cT_n$, $\tilde V$ annihilates $\Sigma(\zeta)$ because $\cY_r$ and hence $\Sigma(\zeta)$ are circle invariant.  Thus the coefficients 
 $A^{(\pm 2)}$ of $\zeta^{\pm 2}$ in $\Sigma(\zeta)$ have weights $\mp 2$ under $V$:
\begin{equation}\label{weight2}
\mathsterling_VA^{(\pm 2)} = \mp 2i A^{(\pm2)}, 
\end{equation}
where $\mathsterling_V A^{(\pm 2)}  = \d\left( V\lrcorner A^{(\pm 2)}  \right)$.  Adding both signs of \eqref{weight2} gives
\begin{equation}
\frac{\partial^2 R}{\partial  x_r^+ \partial x_s^-} = \frac{\partial^2 J}{\partial x_r^+ \partial x_s^-} \,.
\end{equation}
Since $J$ is defined only up to a sum of a function of $x_r^+$ and another of $x_r^-$, we can fix this freedom by identifying $J \equiv R$. Then the difference between the $\pm$ parts of \eqref{weight2} gives
\begin{align}
%[dx_r^+\wedge\d x_s^-] & \qquad 
0 = \partial_{x_r^+} (J^{ts} V_t^+) + \partial_{x_s^-} (J^{rt} V_t^-),\qquad
%[dx_r^+\wedge\d x_s^+] & \qquad  
\epsilon_{rs} =  \partial_{x_s^\pm} ( J^{rt} V_t^\mp),
%\\
%[dx_r^-\wedge\d x_s^-] & \qquad \epsilon_{rs} =  \partial_{x_r^-} ( J^{ts} V_t^+ ).
\end{align}
With $J=R$, the first of these equations reads
\begin{equation}
0 = \partial_{x_r^+} \left( \frac{\partial^2 R}{\partial x_t^+ \partial x_s^-} \frac{\partial R}{\partial x_u^+} \epsilon^{tu} \right) + \partial_{x_s^-} \left( \frac{\partial^2 R}{\partial x_r^+ \partial x_t^-} \frac{\partial R}{\partial x_u^-} \epsilon^{tu} \right). \label{circle0}
\end{equation}
%but this is solved by the constraints
%\begin{equation}
%0 = i_V i_V \Sigma(1) = \frac{\partial R}{\partial x_r^+}\frac{\partial R}{\partial x_s^+} \epsilon^{rs},\qquad 0 = i_V i_V \Sigma(i) = \frac{\partial R}{\partial x_r^-}\frac{\partial R}{\partial x_s^-} \epsilon^{rs}.
%\end{equation}
Likewise, with $J=R$, the remaining equations simplify, and are solved by \eqref{eq:J}, which is now:
\begin{equation}
0= \epsilon_{rs} R^{rr'} R^{ss'} + \epsilon^{r's'}\,.\label{Pleb-R}
\end{equation}
Together, \eqref{circle0} and \eqref{Pleb-R} are an integrable system for $R$. In Lax form, the system is $\{\cL_r, V+i\zeta\p_\zeta\}$, where $V+i\zeta\p_\zeta$ is the circle symmetry generator.

To complete the proof, we show that \eqref{circle0} is linearly independent of \eqref{Pleb-R}. Write $\p^r=\p/\p x^+_r$ and $\p^{r'}=\p/\p x^-_{r'}$. Then \eqref{Pleb-R} can be written as 
\begin{equation}
\p^{[r}( R^{s]s'} R^{r'}\epsilon_{r's'} - \epsilon^{s]q}x^+_q)=0,\qquad\text{or,}\qquad \p^{[r'}( R^{s']s} R^{r}\epsilon_{rs} - \epsilon^{s']q'}x^-_{q'}) = 0\, .\label{R-S-equs}
\end{equation}
These are integrability conditions for the existence of functions $S$ and $S'$ satisfying
\begin{equation}
R^{ss'} R^{r'}\epsilon_{r's'} - \epsilon^{sq}x^+_q=\p^s S\, , \qquad  R^{s's} R^{r}\epsilon_{rs} -\epsilon^{s'q'}x^-_{q'}=\p^{s'} S' \label{circle1}
\end{equation}
where $S$ is defined up to functions of $x^-_{r'}$, and $S'$ is defined up to functions of $x^+_r$. Equation \eqref{circle0} becomes  $\p^r \p^{r'} S' + \p^{r'}\p^r S=0$, which imposes one additional constraint on the system: $S+S'=0$.  \qed
\smallskip 

Note that \eqref{circle1} together with $S+S'=0$,  provides an alternative form of the integrable system, with one fewer derivatives, at the price of introducing the additional function $S$.  %The equations have the trivial flat solution $R=\epsilon^{rs}x^+_r x^-_s$, $S=0$ but it is clear from the expansions arising from the TBA that the true solution is much more complicated.
\section{Discussion}\label{discuss}

The remainder function, $R$, is the key observable for SYM amplitudes. We find that at strong coupling $R$ satisfies an integrable system (for $n$ odd), analogous to the first Plebanski equation for 4d self-dual gravity. Our result follows from a new perspective on the Y-systems of \cite{Alday:2010ku}: they define twistor spaces for the kinematic space, $\cK_n$, of null polygonal Wilson loops with $2n$ sides. Moreover, we find that $\cK_n$ carries a pseudo-hyperk\"ahler structure, for which $R$ is the pseudo-Kahler scalar. These results establish a new geometry underpinning the structure of  Wilson loops  at strong coupling whose study, following the strategy of the conjectures of \cite{Basso:2020xts,Basso:2022ruw} should give insights into the amplitudes to all orders.  Similar ideas apply to other SYM operators, see for example \cite{Bargheer:2019exp,Belitsky:2020qrm} for examples where the imprint from strong coupling can be seen in the full nonperturbative correlator.

We stress that the new pseudo-hyperk\"ahler spaces studied here are \emph{not} related to other hyperk\"ahler structures that arise in studies of Hitchin systems. Hitchin showed that moduli spaces of regular Hitchin systems admit hyperk\"ahler structures, \cite{Hitchin:1988df} and this result has been partially extended to the irregular case in  \cite{Biquard:2004} and more analogously by  Gaiotto, Moore and Neitzke  in \cite{Gaiotto:2008cd,Gaiotto:2009hg}. However, neither of these apply directly to amplitudes at strong coupling, because $\cK_n$ parameterises an \emph{invariant subspace} of the Hitchin moduli space under an involution \cite{Alday:2009yn,Alday:2009dv,Alday:2010vh}.  We have verified that the standard hyperk\"ahler structures do not restrict to this subspace to yield our results, even in simple examples.

We comment on implications of our results.
First, these methods  apply  to other cluster varieties/$Y$-systems, such as the ADE-type cases \cite{Zamolodchikov:1991et}, and the affine and surface-type cluster algebras. Physically, type $D$ corresponds to  form factors at strong coupling for restricted kinematics. Beyond these cases, Grassmannian cluster algebras appear when computing strong coupling amplitudes and form factors for full $\mathcal{N}=4$ SYM kinematics. The Y-systems associated to these cluster algebras are known, \cite{Alday:2010vh} as are generalizations  incorporating form-factors  \cite{Maldacena:2010kp,Gao:2013dza}. Our strategy developed here will lead to integrable systems for the amplitudes in all of these cases, albeit with novelties arising beyond restricted kinematics. %$\cK_n$ for the full $\mathcal{N}=4$ strong coupling amplitude has dimension $3n-15$, and so can only be hyperk\"ahler for $n=9,13,17...$.

It is well known that soft and multi-colinear limits provide a system of boundary conditions on the remainder function, see \cite{Alday:2009yn, Maldacena:2010kp}, and in particular equation (2.10) of \cite{Goddard:2012cx} which gives the boundary condition
\begin{equation}
    R_n\rightarrow R_{n-m}+R_{m+4}\, ,
\end{equation}
see also \cite{Caron-Huot:2020bkp} for a recent summary at weak coupling. Coupled with the differential equations we have found here, this will lead to a unique determination of the remainder function starting with the smallest nontrivial boundary conditions provided by the octagon, which is treated in full detail in \cite{Alday:2009yn}: this clearly shows that the full solution is highly nontrivial with a nontrivial infinite series beyond the trivial  quadratic solution $R=\epsilon_{rs}x^r_+x^s_-$.

% The colinear limit, for two adjacent vertices of the Wilson loop, defines a codimension two boundary of $\cK_n$ that can be identified with $\cK_{n-1}$.  On these boundaries, $R$ can be simply determined in terms of the $\cK_{n-1}$ remainder functions (or at least the output of the $Y$-system) and other known quantities 
%Although these boundaries have co-dimension two, our equations are over-determined by at least that co-dimension.  %For example, $\cK_5$ is 4-dimensional. %The Plebanski equations are deterministic in 4d, but subject to the circle symmetry, they should propagate data from the 2d $\cK_4$ boundaries (where $R$ is known \cite{Alday:2009yn}).
%Our equations become more over-determined as $n$ grows as   $\cT_n$ has complex dimension $n-2$, whereas $\cK_n$ has dimension $2n-6$. Thus these equations should give an inductive scheme to solve $R$ at all $n$.  

%As an aside, note that the area is positive on the Euclidean region, in which non-adjacent vertices are space-like separated. So imposing positivity on the remainder function $R$ may be a useful constraint to use when solving for $R$.

Finally, our results suggest avenues beyond the strong coupling limit. The work of \cite{Basso:2020xts,Basso:2022ruw} identifies lines in the full kinematic space where  quadratic log solutions are valid at strong coupling, and explains how formulae for all values of the coupling may be obtianed from the the solutions to the $Y$-system there may be found.   Thus these structures are likely to be important for extensions to these conjectures. 
 Our differential equations at strong coupling, and associated structures might therefore  be deformable to some that hold beyond strong coupling. In this direction, there are several other connections to explore. Our integrable system can be recovered from a twistor sigma model action \cite{Adamo:2021bej,Mason:2022hly}; quantizing analogous  models might allow computations beyond the strong coupling limit. 
Related structures arise for the anomalous dimension spectrum at finite coupling in the form of the Y-system of the quantum spectral curve, \cite{Gromov:2014caa} with the coupling constant  incorporated via the `Joukowski correspondence', \cite{Mason:1988w,Mason:1996w, Ferrari:2018ubt}.

\paragraph{Acknowledgements:} It is a pleasure to acknowledge 
conversations with Benjamin Basso, Philip Boalch, Tom Bridgeland, Lance Dixon,  Volodya Kazakov and Fedor  Levkovich-Maslyuk. LJM would also like to thank the Institutes des Haut \'Etudes Scientifique, Bures Sur Yvette, and the Laboratoire Physique at the ENS, Paris  for hospitality while this was being written up, the STFC for financial support from  grant numbers ST/T000864/1, ST/X000761/1 and the Simons Collaboration on Celestial Holography.  \"OCG is supported by UKRI/EPSRC Stephen Hawking Fellowship EP/T016396/1 and the Royal Society University
Research Fellowship URF\textbackslash R1\textbackslash 221236. HF is supported by a Merton College Junior Research Fellowship.

\bibliographystyle{JHEP}
\bibliography{sdpw1}

\end{document}

%% file: figures.tex
\def\segmentup[#1,#2,#3,#4]{

    \pgfmathsetmacro{\tini}{pi*#1}
    \pgfmathsetmacro{\tnext}{pi*#1+\step}
    \pgfmathsetmacro{\tfin}{pi*#2}

  \foreach \angulo in {\tini,\tnext,...,\tfin}{
    \pgfmathsetmacro{\radius}{3}

    \pgfmathparse{\angulo+\step}
    \pgfmathsetmacro{\angulonext}{\pgfmathresult}
    \pgfmathparse{\angulo+2*\step}
    \pgfmathsetmacro{\angulonnext}{\pgfmathresult}

    \pgfmathparse{\radius*cos(\angulo r)}
    \pgfmathsetmacro{\px}{\pgfmathresult}
    \pgfmathparse{\radius*sin(\angulo r)}
    \pgfmathsetmacro{\py}{\pgfmathresult}
    \pgfmathparse{\radius*pi*#3+\radius*(\angulo)-\radius*(\tini)}
    \pgfmathsetmacro{\pz}{\pgfmathresult}

    \pgfmathparse{\radius*cos(\angulonext r)}
    \pgfmathsetmacro{\pxnext}{\pgfmathresult}
    \pgfmathparse{\radius*sin(\angulonext r)}
    \pgfmathsetmacro{\pynext}{\pgfmathresult}
    \pgfmathparse{\radius*pi*#3+\radius*(\angulonext)-\radius*(\tini)}
    \pgfmathsetmacro{\pznext}{\pgfmathresult}

    \draw[#4,postaction={decorate}] (\px,\py,\pz) -- (\pxnext,\pynext,\pznext);

   \draw[draw=none,top color=red,bottom color=white] (\px,\py,\pz-0.8) rectangle (\pxnext,\pynext,\pznext);

  }
}

\def\segmentdown[#1,#2,#3,#4]{

    \pgfmathsetmacro{\tini}{pi*#1}
    \pgfmathsetmacro{\tnext}{pi*#1+\step}
    \pgfmathsetmacro{\tfin}{pi*#2}

  \foreach \angulo in {\tini,\tnext,...,\tfin}{
    \pgfmathsetmacro{\radius}{3}

    \pgfmathparse{\angulo+\step}
    \pgfmathsetmacro{\angulonext}{\pgfmathresult}
    \pgfmathparse{\angulo+2*\step}
    \pgfmathsetmacro{\angulonnext}{\pgfmathresult}

    \pgfmathparse{\radius*cos(\angulo r)}
    \pgfmathsetmacro{\px}{\pgfmathresult}
    \pgfmathparse{\radius*sin(\angulo r)}
    \pgfmathsetmacro{\py}{\pgfmathresult}
    \pgfmathparse{\radius*pi*#3-\radius*(\angulo)+\radius*(\tini)}
    \pgfmathsetmacro{\pz}{\pgfmathresult}

    \pgfmathparse{\radius*cos(\angulonext r)}
    \pgfmathsetmacro{\pxnext}{\pgfmathresult}
    \pgfmathparse{\radius*sin(\angulonext r)}
    \pgfmathsetmacro{\pynext}{\pgfmathresult}
    \pgfmathparse{\radius*pi*#3-\radius*(\angulonext)+\radius*(\tini)}
    \pgfmathsetmacro{\pznext}{\pgfmathresult}

    \draw[#4,postaction={decorate}] (\px,\py,\pz) -- (\pxnext,\pynext,\pznext);
    \draw[draw=none,top color=white!50!red,bottom color=white] (\px,\py,\pz-0.8) rectangle (\pxnext,\pynext,\pznext);
  }
}

\def\segmentupnew[#1,#2,#3]{
  \pgfmathparse{#2+#3}
  \pgfmathsetmacro{\thmax}{\pgfmathresult}
  \pgfmathparse{#2}
  \pgfmathsetmacro{\thmin}{\pgfmathresult}
  \pgfmathparse{#1}
  \pgfmathsetmacro{\zzero}{\pgfmathresult}

  \draw[domain=\thmin:\thmax,smooth,variable=\t] plot (
  {3*cos(\t r)},
  {3*sin(\t r)},
  {\zzero + 3*(\t - \thmin)});
}

\def\segmentupnewdashed[#1,#2,#3]{
  \pgfmathparse{#2+#3}
  \pgfmathsetmacro{\thmax}{\pgfmathresult}
  \pgfmathparse{#2}
  \pgfmathsetmacro{\thmin}{\pgfmathresult}
  \pgfmathparse{#1}
  \pgfmathsetmacro{\zzero}{\pgfmathresult}

  \draw[dashed, domain=\thmin:\thmax,smooth,variable=\t] plot (
  {3*cos(\t r)},
  {3*sin(\t r)},
  {\zzero + 3*(\t - \thmin)});
}

\def\segmentdownnew[#1,#2,#3]{
  \pgfmathparse{#2+#3}
  \pgfmathsetmacro{\thmax}{\pgfmathresult}
  \pgfmathparse{#2}
  \pgfmathsetmacro{\thmin}{\pgfmathresult}
  \pgfmathparse{#1}
  \pgfmathsetmacro{\zzero}{\pgfmathresult}

  \draw[domain=\thmin:\thmax,smooth,variable=\t] plot (
  {3*cos(\t r)},
  {3*sin(\t r)},
  {\zzero - 3*(\t - \thmin)});
}

\def\segmentdownnewdashed[#1,#2,#3]{
  \pgfmathparse{#2+#3}
  \pgfmathsetmacro{\thmax}{\pgfmathresult}
  \pgfmathparse{#2}
  \pgfmathsetmacro{\thmin}{\pgfmathresult}
  \pgfmathparse{#1}
  \pgfmathsetmacro{\zzero}{\pgfmathresult}

  \draw[dashed, domain=\thmin:\thmax,smooth,variable=\t] plot (
  {3*cos(\t r)},
  {3*sin(\t r)},
  {\zzero - 3*(\t - \thmin)});
}

\tikzstyle{quiver}=[
thin, -latex, shorten >= 0.5mm,blue
]
\tikzstyle{chord}=[
    thick
    ]
    \tikzstyle{chordmiddle}=[
    inner sep=0,outer sep=0
    ]

\newcommand{\zigzagcluster}{
  \begin{tikzpicture}
                    \draw[white] (-3,-2) rectangle (3,3);
   \node (s)   [regular polygon, regular polygon sides=9, draw=none, minimum size=33mm, shape border rotate=40] {};

	\node[above] (11) at (s.corner 1) {1};
		\node[above] (12) at (s.corner 9) {2};
		\node[left] (10) at (s.corner 2)	 {$n$};
		\node[left] (14) at (s.corner 3)	 {$n-1$};
                \node[below left] (15) at (s.corner 4)	 {$n-2$};
                \node[above right] (15) at (s.corner 8) {3};
                \foreach \i in {1,2,3,0,-1,-2,4,5}{
                  \pgfmathparse{int((\i) - 9*floor((\i-1)/9)))};
                  \pgfmathsetmacro{\from}{\pgfmathresult};
                  \pgfmathparse{int((\i + 1) - 9*floor((\i)/9)))};
                  \pgfmathsetmacro{\to}{\pgfmathresult};
                  \draw[chord] (s.corner \from) -- (s.corner \to);
                }
                \draw[chord] (s.corner 1) -- (s.corner 3);
                \draw[chord] (s.corner 1) -- (s.corner 4);
                \draw[chord] (s.corner 4) -- (s.corner 9);
                \draw[chord] (s.corner 9) -- (s.corner 5);
                \draw[chord] (s.corner 5) -- (s.corner 8);
                \draw[dotted] (s.corner 6) -- (s.corner 7);

                \node[chordmiddle] (b1n) at ($(s.corner 1)!0.5!(s.corner 2)$) {};
                \node[chordmiddle] (b12) at ($(s.corner 9)!0.5!(s.corner 1)$) {};
                \node[chordmiddle] (b23) at ($(s.corner 8)!0.5!(s.corner 9)$) {};
                \node[chordmiddle] (b34) at ($(s.corner 7)!0.5!(s.corner 8)$) {};
                \node[chordmiddle] (bdots) at ($(s.corner 6)!0.5!(s.corner 5)$) {};
                \node[chordmiddle] (bnnm1) at ($(s.corner 2)!0.5!(s.corner 3)$) {};
                \node[chordmiddle] (bnm1nm2) at ($(s.corner 3)!0.5!(s.corner 4)$) {};
                \node[chordmiddle] (bnm2nm3) at ($(s.corner 4)!0.5!(s.corner 5)$) {};
                \node[chordmiddle] (bnm3nm4) at ($(s.corner 5)!0.5!(s.corner 6)$) {};

                \node[chordmiddle] (b1nm1) at ($(s.corner 1)!0.5!(s.corner 3)$) {};
                \node[chordmiddle] (b1nm2) at ($(s.corner 1)!0.5!(s.corner 4)$) {};
                \node[chordmiddle] (bnm22) at ($(s.corner 4)!0.5!(s.corner 9)$) {};
                \node[chordmiddle] (b2nm3) at ($(s.corner 9)!0.5!(s.corner 5)$) {};
                \node[chordmiddle] (bnm33) at ($(s.corner 5)!0.5!(s.corner 8)$) {};

                \draw[quiver]  (b1n) -- (b1nm1);
                \draw[quiver]  (b1nm1) -- (bnnm1);
                \draw[quiver]  (bnnm1) -- (b1n);
                \draw[quiver]  (b1nm1) -- (b1nm2);
                \draw[quiver]  (bnm1nm2) -- ( b1nm1)  ;
                \draw[quiver]  (b12)  -- (bnm22)  ;
                \draw[quiver]  (b1nm2) -- (b12) ;
                \draw[quiver]  (b1nm2) -- (bnm1nm2)  ;
                \draw[quiver]  (bnm22)  -- (b1nm2)  ;
                \draw[quiver]  (bnm22)  -- (b2nm3)  ;
                \draw[quiver]  (b2nm3) -- (bnm2nm3) ;
                \draw[quiver]  (bnm2nm3) -- (bnm22) ;

                \draw[quiver]  (b2nm3) -- (b23) ;
                \draw[quiver]  (b23) -- (bnm33) ;
                \draw[quiver]  (bnm33) -- (b2nm3) ;
                
                \draw[quiver]   (bnm3nm4) -- (bnm33) ;
                \draw[blue, thin, dotted]    (bnm33)  -- ($(0.4,0)+ (bnm33)$);

              \end{tikzpicture}
            }

            \newcommand{\exxodd}{
              \tikz{
                \draw[white] (-1,-2) rectangle (3,3);
              \coordinate (i) at (2,1.5) ;
              \coordinate (im1) at (0,1.5) ;
              \coordinate (j) at (2,-0.5) ;
              \coordinate (jp1) at (0,-0.5) ;

              \draw[chord] (i) node[above] {$j$} -- (im1) node[above] {$i$};
              \draw[chord] (j)  node[below] {$k$}-- (jp1) node[below] {$l$};
              \draw[chord] (i) -- (jp1);
              \draw[chord] (im1) -- (jp1);
              \draw[chord] (i) -- (j);

              \draw[quiver] ($(im1)!0.5!(i)$) -- ($(im1)!0.5!(j)$);
              \draw[quiver] ($(j)!0.5!(jp1)$) -- ($(im1)!0.5!(j)$);

              \draw[quiver]  ($(im1)!0.5!(j)$) -- ($(j)!0.5!(i)$);
              \draw[quiver]  ($(im1)!0.5!(j)$) -- ($(jp1)!0.5!(im1)$) ;

              \node (rel) at (1,-1.2) {\rotatebox{90}{$\leftrightarrow$}};
              \node (xr) at (1,-2) {$(i\,l|k\,j)^\pm$};

            }
            }

\newcommand{\zigzaganan}{
  \begin{tikzpicture}[scale=0.7]
    \draw[-latex] (-0.5,0.5) -- (5,-5) node[right] (xp) {$X^+$};
    \draw[-latex] (-0.5,-0.5) -- (5,5) node[left] (xm) {$X^-$};
    \draw[thick] (0,0) -- (1,-1) -- ($(1,-1)+(1.2,1.2)$) -- ($(1.6,-1.6)+(1.2,1.2)$) -- ($(1.6,-1.6)+(2.1,2.1)$) -- ($(2.2,-2.2)+(2.1,2.1)$) -- ($(2.2,-2.2)+(3,3)$);
    \draw[thick, dotted] ($(2.2,-2.2)+(3,3)$) -- ($(2.5,-2.5)+(3,3)$);
    \draw[dotted] ($(1.6,-1.6)+(1.2,1.2)$) -- ($(1.6,-1.6)+(0,0)$) node[below left] (xi) {$\scriptstyle X^+_i$} ; 
    \draw[dotted] ($(2.2,-2.2)+(2.1,2.1)$) -- ($(2.2,-2.2)+(0,0)$) node[below left] (xi) {$\scriptstyle X^+_{i+1}$} ;
    \draw[dotted] ($(1,-1)+(1.2,1.2)$) -- ($(0,0)+(1.2,1.2)$) node[above left] (xi) {$\scriptstyle X^-_{i}$} ;
    \draw[dotted] ($(1.6,-1.6)+(2.1,2.1)$) -- ($(0,0)+(2.1,2.1)$) node[above left] (xi) {$\scriptstyle X^-_{i+1}$} ;
    \draw [gray, decorate,decoration={brace,amplitude=5pt,mirror,raise=6ex}]
  (0,0) -- (5,-5) node[midway,yshift=-3em,xshift=-3em]{$\cM^\R_{0,n}$};
    \draw [gray,decorate,decoration={brace,amplitude=5pt,mirror,raise=6ex}]
  (5,5) -- (0,0) node[midway,yshift=3em,xshift=-3em]{$\cM^\R_{0,n}$};
  \end{tikzpicture}
}